\input psfig
\documentstyle[twocolumn,aps]{revtex}
\begin{document}
\draft
\twocolumn[\hsize\textwidth\columnwidth\hsize\csname @twocolumnfalse\endcsname
%
%

\title{ A Model of Correlated Fermions with $d_{x^2-y^2}$ Superconductivity}

\author{ Alexander Nazarenko$^a$, Adriana Moreo$^a$,
Jose Riera$^b$, and Elbio Dagotto$^a$}

\address{$a.$ Department of Physics, and National High Magnetic Field Lab,
Florida State University, Tallahassee, FL 32306, USA}
\address{$b.$ Instituto de Fisica Rosario, Avda. 27 de Febrero 210 bis,
2000 Rosario, Argentina}

\date{\today}
\maketitle

\begin{abstract}
Motivated by the phenomenology of the high-Tc cuprates,
a two dimensional fermionic model with attractive interactions
is here discussed.
The exact solution to the two particle
problem leads to a bound state in the $d_{x^2 - y^2}$ subspace.
Numerical techniques suggest that the model has $d_{x^2 - y^2}$
superconductivity (SC) in the ground state at low fermionic density.
Within a self-consistent RPA diagrammatic study, the density dependence of
the critical temperature is calculated. We argue
that in the context of d-wave SC  this model fulfills
the role that the attractive on-site Hubbard model has played for
s-wave SC. We also show that another candidate, the
attractive ``t-U-V'' model, which has d-wave SC at the mean-field level
is actually not useful as a realization of this family of condensates
for a variety of reasons.

\end{abstract}

\pacs{74.20.-z, 74.20.Mn, 74.25.Dw}
\vskip2pc]
\narrowtext

%
%

The Hubbard model with an $attractive$ on-site
interaction has played an important role in the
qualitative understanding of s-wave superconductors.
Several interesting problems, like
the crossover from the
BCS regime to the region where Copper pairs form a
Bose Condensate (BC),\cite{leggett}
can be addressed within this model using analytical and numerical
techniques.\cite{scalettar}
Although the phononic degrees of freedom are not explicit, it
is expected that the qualitative properties of the attractive Hubbard
model are the same as those of more realistic, and difficult to study,
electron-phonon Hamiltonians.

However, since evidence is accumulating that the high critical temperature
(high-Tc) superconductors have a condensate with pairs formed in the
${ d_{x^2 - y^2}}$ channel,\cite{doug} the relevance for the cuprates
of the on-site attractive Hubbard
model is questionable. In addition,  studies of $H_{c2}$ in cuprates
suggest that the Cooper pairs are only
a few lattice spacings in size,
locating the high-Tc materials in an intermediate
region between the BCS and the BC limits. In this regime, the
BCS mean-field (MF) approximation is not quantitatively
accurate.\cite{sademelo} Thus, it
would be desirable to have a fermionic model
with a $d_{x^2 - y^2}$
superconducting ground state that can be studied with reliable
non-perturbative
diagrammatic and computational techniques in the small coherence length
region. Results obtained in this framework could be directly compared with
the cuprate phenomenology.

What model can provide the generalization of the attractive
Hubbard model for the case of  $d_{x^2-y^2}$ superconductivity?
Current literature shows that many studies
of d-wave superconductors are performed
using the BCS MF approximation after
introducing a proper attractive
kernel in the gap equation to induce d-wave correlations.
This approach is accurate in weak coupling where pairs are large,
but it does not address the regime of small pairs which
is more realistic for the cuprates.
To improve these results, it is natural to
consider as a first candidate for a model with d-wave SC the
so-called
``t-U-V'' model\cite{micnas} where U is repulsive on-site, V is the strength
of a
density-density attraction at distance of one lattice spacing $a$, and t
is the amplitude of a nearest-neighbors (n.n.) hopping term.
In the MF approximation\cite{micnas}
the phase diagram at half-filling of the t-U-V model (Fig.1a)
indeed has an ``island'' of d-wave SC.
However, note
the small size of this phase caused by the
competition with phase separation (PS) where electrons doubly occupy a
macroscopic region of the cluster to minimize the energy. A recent
study\cite{haas} has shown that the effect of PS cannot be simply
avoided by introducing a long-range Coulomb
repulsion since in this situation PS may be
replaced by a charge-density-wave (CDW) state
rather than by SC. Thus, the competition SC-PS is subtle and not
created by the absence of long-range interactions.

Using techniques more powerful than MF,
in this paper we argue that the t-U-V model is not much useful
for the study of d-wave superconductors.
As a first indication of this problem, let us analyze the results obtained
using Quantum Monte Carlo (QMC) and Exact Diagonalization (ED)
techniques.\cite{review} Working at small U/t
where the MF approximation predicts a
d-wave condensate, first we have studied the competition with PS.
This phase can be observed numerically with QMC
simulations since in some  regions of parameter space we
found that the mean particle density
converges to two very different results depending on the
randomnly chosen initial
Hubbard-Stratonovich fields. Such a behavior is
typical of systems with two competing minima in the free energy, as it
happens in the presence of PS. In QMC the effect can already be seen
even using small clusters of 16 and 36 sites, and working at relatively
high temperatures $T=t/6$ to alleviate sign
problems. As example, the presence of the two
minima was numerically observed at $U/t=1$ for $|V/t| > 0.55$,
reducing the MF d-wave stability
region (Fig.1a).
Increasing the lattice size and reducing the temperature would
only enhance the effect, since the
tunneling probability would become negligible.
\begin{figure}[htbp]
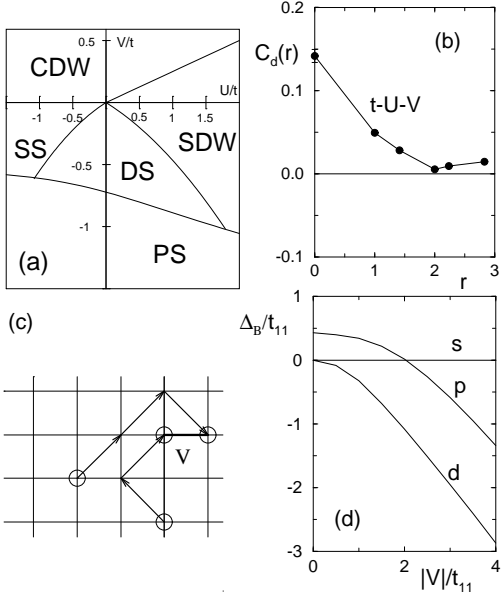

\centerline{\psfig{figure=Fig1_a_dw.epsi,height=3.45cm,angle=270}
\hspace{.1cm}\psfig{figure=Fig1_b_dw.epsi,height=3.81cm,angle=270}}
\centerline{\psfig{figure=Fig1_c_dw.epsi,height=3.cm,angle=90}
\hspace{.105cm}\psfig{figure=Fig1_d_dw.epsi,height=4cm,angle=270}}
\vspace{.2cm}
\caption{(a) Mean-field phase diagram of the ``t-U-V'' model at
half-filling.[5] DS,SS,PS,CDW, and SDW denote d-wave
SC, s-wave SC, phase separation,
charge-density-wave,  and spin-density-wave order,
respectively;
(b) ${\rm d_{x^2 - y^2}}$ pairing correlation vs
distance $r$ obtained with
QMC, on a $4 \times 4 $ cluster, ${ T=t/6}$,
${ U/t=0.0}$ and ${V/t=-0.3}$ i.e. within the MF
d-wave region. The correlations are small at distances
larger than one. Similar negative results were obtained for other couplings
inside the MF d-wave region. The error bars are smaller than the
dots;
(c) Schematic representation of model Eq.(1). Fermions move within the same
sublattice, and interact at distance one; (d) Binding energy $\Delta_B$
for the two-particle problem solved exactly using $t_{20} = 0.4 t_{11}$, as
example. States with the lowest energy in the $d_{x^2 - y^2}$,
extended-$s$ and $p$-subspaces are
shown. Similar results are found for $t_{20} < 0.5 t_{11}$.
If $t_{20} > 0.5 t_{11}$, at small $|V|/t_{11}$ the actual ground
state has $p$-wave, but it becomes again
$d_{x^2-y^2}$-wave after a finite coupling is reached.}
\end{figure}

In addition to this problem, in the regime where
PS is not observed in our analysis,
a QMC simulation still does not show enhanced d-wave correlations at
$T=t/6$. Defining the operator that destroys a d-wave local pair
as ${ \Delta^d_{\bf i} = c_{{\bf i}\uparrow} ( c_{{\bf i+{\hat
x}}\downarrow}
+ c_{{\bf
i-{\hat x}}\downarrow} - c_{{\bf i+{\hat y}}\downarrow} -
c_{{\bf i-{\hat y}}\downarrow} ) }$, with ${\bf
{\hat x},{\hat y}}$ unit vectors along the axis, then the ground state
pairing correlation is
${ C_{d}({\bf r}) = \langle {\Delta^d_{\bf i}}^\dagger \Delta^d_{\bf i+r}
\rangle }$.  Fig.1b shows that $C_d({\bf r})$
decays to zero with distance. Actually, results at $U,V=0$ are very
similar to Fig.1b. We also performed
ED studies on $4\times 4$ clusters that
are in excellent agreement with the QMC data.
While the QMC result does not rule
out the stability of d-wave SC at $T \ll t/6$, it
shows that (i) the small region of possible d-wave SC cannot be easily
studied with
current state-of-the-art computational
studies beyond
the MF approximation, and (ii) obviously in such a weakly correlated
regime, Cooper pairs can only be large in size contrary to the
high-Tc phenomenology. In addition, note also that
the tentative d-wave SC regime of the t-U-V model occurs at
half-filling with all electrons pairing to form the condensate. However,
it would be desirable to have a model where
a $small$ density of carriers forms
pairs, mimicking the expected hole-pairing of
the cuprates. The t-U-V model at low electronic
density has s-wave SC rather
than d-wave,\cite{micnas} complicating matters further.
An inescapable conclusion of this analysis is
that it is necessary to go beyond the t-U-V model for a proper study
of effective fermionic models for d-wave SC.\cite{comment}
It is remarkable that in spite of the recent huge effort devoted
to the study of d-wave
SC in the cuprates, the analog
of the attractive Hubbard model for $d_{x^2 - y^2}$ pairs still seems
unknown.

The main purpose of this paper is to discuss
a fermionic model for $d_{x^2 - y^2}$ SC which solves
the problems found in the t-U-V Hamiltonian. Analyzing
the model studied  here with computational techniques, it
presents strong pairing correlations in the
d-wave channel, and PS does not cause serious problems.
The Hamiltonian contains
an attractive n.n. density-density interaction at distance $a$,
as in the t-U-V model,
but it differs from it in the fermionic dispersion
which in the new model is dominated by hopping within the $same$
sublattice, i.e. linking next-nearest-neighbor (n.n.n.)
sites. This model was discussed before in the context of the
``Antiferromagnetic van Hove''(AFVH)  scenario for the cuprates where
the high $T_c$ is induced  by a large peak in the hole density of
states (DOS) caused by
antiferromagnetism.\cite{afvh} The intra-sublattice
dispersion is natural if holes move in a nearly
antiferromagnetic background that is energetically costly to disturb.
The attractive  interaction has its origin in AF correlations. In
Ref.\cite{afvh} the model was studied only within a MF approximation,
but here we substantially improve the analysis using computational and
self-consistent diagrammatic techniques.
The Hamiltonian of the proposed model candidate for d-wave SC is
\begin{equation}
H = \sum_{{\bf k}\alpha} \epsilon_{AF}({\bf k})
[c^\dagger_{{\bf k}\alpha} c_{{\bf
k}\alpha} + h.c.]
- |V| \sum_{\langle {\bf ij} \rangle } n_{\bf i} n_{\bf j},
\end{equation}
where $\alpha=A,B$ indicates the sublattice;
 $\epsilon_{AF}({\bf k})= 4t_{11} cosk_x cosk_y + 2t_{20}
(cos2k_x + cos2k_y)$ is the dispersion; $t_{11},t_{20},V$ are parameters;
and $n_{\bf i}$ is the number operator, with the
rest of the notation standard (see Fig.1c). The operators $c$
satisfy anticommutation relations. They do not have a spin index, but
carry a sublattice index which plays a similar role. Particles are
distributed such that half of them are in each sublattice.
Intuitively the particles described by
Eq.(1) represent``holes'' in the cuprates.

It is straightforward, and very instructive, to solve exactly the two particle
problem using Eq.(1). Defining the binding
energy as ${\Delta_B = E_2 - 2 E_1}$, where ${E_n}$ is the
ground state energy of the $n$-particles subspace,
the results are shown in Fig.1d for extended-$s$, $d_{x^2 - y^2}$ and
$p$-wave symmetries. The lowest energy state has ${\Delta_B < 0}$,
signaling the presence of a bound state, and it corresponds to $d_{x^2
- y^2}$-symmetry. At small ${ |V|}$, i.e. for weakly bounded
particles, ${\Delta_B}$ is very small in absolute value, but still negative.
Studying the average distance between the two particles
we observed that at ${ |V|/t_{11} \sim 2}$ in Fig.1d, the pair
size is already close to its maximum value of one
lattice spacing defining the strong coupling regime.
\begin{figure}[htbp]
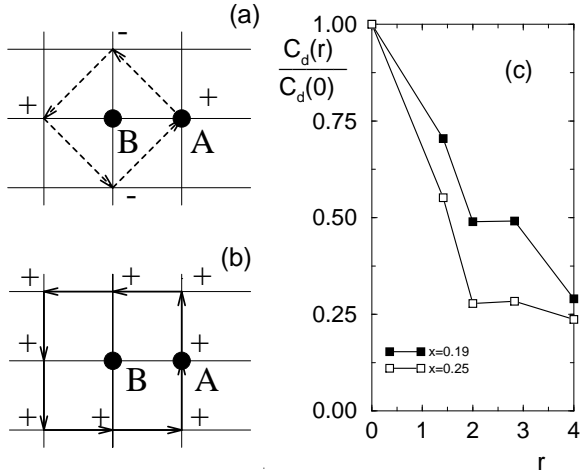

\centerline{\psfig{figure=Fig2_ab_dw.epsi,height=6cm,angle=90}
\hspace{.1cm}\psfig{figure=Fig2_c_dw.epsi,height=6cm,angle=270}}
\vspace{.2cm}
\caption{ (a) Two fermions in the large $|V|$ limit of Eq.(1) showing the
$d_{x^2 - y^2}$-wave character of the bound state; (b) Two fermions in the
large $|V|$ limit of the t-U-V model. The bound state is $s$-wave; (c)
$d_{x^2-y^2}$ pairing correlations vs distance $r$ for model Eq.(1) at T=0
studied with ED techniques on
a 32 site cluster. The couplings are
$|V|/t_{11} = 1.0$, $t_{20}/t_{11} = 0.4$, and $x$ is indicated.
Correlations in the $s-$ and $p$-channels are negligible.}
\end{figure}

What is the origin of the d-wave symmetry in the
ground state of the two-body problem?
Consider the large ${ |V|}$
limit, and let us analyze the movement of one particle around the other as
schematically shown in Fig.2a. In this regime, the energy is minimized
when the interparticle distance is one lattice spacing
at all times. Keeping
particle B fixed at a given site, the problem now amounts to
solving an effective four-site hopping Hamiltonian of particle A moving along
the four n.n. sites to B using the hoping amplitude $t_{11}$
along the diagonal. Here it is important to observe
that the sign of ${ t_{11}}$ is chosen as a $positive$ number
by the requirement that the minimum in the dispersion is at ${\bf p}
=(\pi/2,\pi/2)$ or $(0,\pi)$, as it is natural in
problems of holes in antiferromagnets.\cite{afvh} The signs of
$t_{11}$ and $t_{20}$ are
physically relevant, unlike the sign of a n.n. hopping
that can be changed by suitable transformations on a square lattice.
Then, the ground state of the effective four-site
problem  corresponds to selecting
a phase alternating in sign for
particle A (Fig.2a). This leads to a $d_{x^2 - y^2}$ bound state,
providing a real-space intuitive explanation for the appearance of
d-wave pairs which complements those based on the perturbative
interchange of magnons.\cite{doug}
We remark that this simple result found in Eq.(1)
is not present in the t-U-V model.
If the n.n.n. hopping  (Fig.2a) is replaced by the
n.n. hopping of the t-U-V model, then the ground state phases of
particle A orbiting around
B at large $|V|$ are as shown in Fig.2b. They correspond
to an s-wave bound state.

The presence of $d_{x^2 - y^2}$
bound states in the two body problem of Eq.(1) suggests
SC in the same channel at finite particle
density. However,
CDW and PS states are also favored by a particle-particle attraction
and thus an explicit calculation is needed to verify the existence of a
SC condensate.
For this purpose, we exactly calculated the ground-state pairing
correlations $C_d({\bf r})$ on a $\sqrt{32}\times \sqrt{32}$  site
cluster using ED techniques. The operator is the same used before for
the t-U-V model, simply switching the spin indices for sublattice indices.
In order to avoid possible complications with the
potentially dangerous PS regime, in the study shown
in Fig.2c an intrasublattice
density-density repulsive interaction of strength $|V|/\sqrt{2}$ and
$|V|/2$ at distances
$\sqrt{2}a$ and $2a$, respectively, was also included in the Hamiltonian.
The results at ${ |V/t_{11}| = 1.0}$, $t_{20} = 0.4 t_{11}$, and at
realistic densities $x=6/32 \sim 0.19$ and $8/32 = 0.25$,
are shown
in Fig.2c. The robust tail suggests
strong pairing correlations in the ground state.
Thus, with Hamiltonian Eq.(1) supplemented by
mild assumptions to avoid PS it is possible to
obtain numerical signals of d-wave
SC, unlike the results obtained before for the
t-U-V model.\cite{comm4,comm3}

Hamiltonian Eq.(1) can also be studied diagrammatically using
Eliashberg-type equations.\cite{comm7} With this
approach, we have calculated $T_c$ vs $x$ in the
intermediate to strong coupling region.
Working in Matsubara space and in natural units,
we approximate the normal state proper self-energy
by iteratively solving the equation
\begin{equation}
\Sigma({\bf k},\omega_n) =-\frac{T}{N}\sum_{{\bf q},{\omega_{n}^{\prime}}}
V_{eff}({\bf k}-{\bf q}, \omega_{n}-
\omega_{n}^{\prime})G({\bf q}, \omega_{n}^{\prime}),
\end{equation}
where $N$ is the number of sites (we used a $32 \times 32$ cluster for
this calculation),  $T$ the temperature,
$\omega_n = (2n+1) \pi T$ ($-\infty < n < +\infty$),
the full normal state one particle Green's function satisfies
$ G({\bf k}, \omega_{n})= 1/
[ {i\omega_n-(\epsilon_{AF}({\bf k})-\mu)-\Sigma({\bf k}, \omega_n)} ]$,
and $V_{eff}({\bf k}, \omega_{n})$ is the RPA effective potential
with particle-hole bubbles containing $G$ rather than the
noninteracting Green's function, to make the calculation
self-consistent.
Once the normal state G is found, for the SC state we use
\begin{equation}
\Phi({\bf k},\omega_n) = \sum_{{\bf q},{\omega_{n}^{\prime}} }
M({\bf k},{\bf q},\omega_n, {\omega_{n}^{\prime}})
\Phi({\bf q},\omega_{n}^{\prime}),
\end{equation}
where $\Phi({\bf k},\omega_n)$ is the anomalous self-energy which can be
considered as an order parameter for SC, and
$M({\bf k},{\bf
q},\omega_n,{\omega_{n}^{\prime}})=-{\frac{T}{N}}V_{eff}({\bf k-q},
\omega_{n}-\omega_{n}^{\prime})$ $G({\bf q}, \omega_{n}^{\prime})
G(-{\bf q}, -\omega_{n}^{\prime})$.
We have solved Eqs.(2,3) self-consistently at different
temperatures and densities, using as an energy cutoff ten times the bandwidth.
The symmetry of the SC condensate is determined from
the symmetry of the eigenvector of Eq.(3) corresponding
to the largest eigenvalue.\cite{comm7}
After all ring diagrams
are summed up,\cite{comm5} this symmetry is $d_{x^2-y^2}$.
\begin{figure}[htbp]
\centerline{\psfig{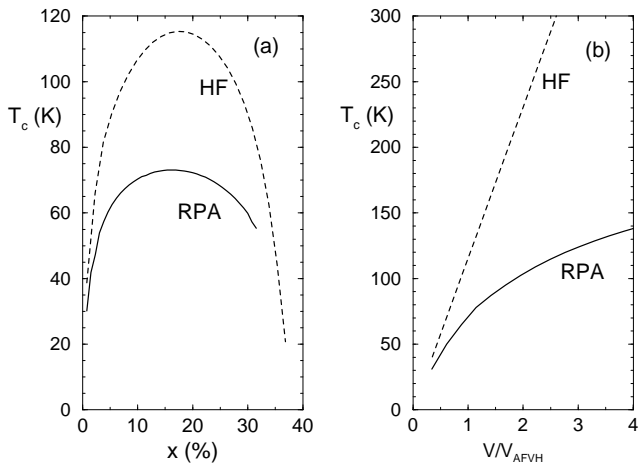}}
\vspace{.2cm}
\caption{(a) Critical temperature $T_c$ of model Eq.(1) vs
fermionic density $x$, using $t_{11} = 0.165 eV$, $t_{20} = 0.0435 eV$, and
$V= -0.075 eV$, as suggested by the high-Tc phenomenology.[9]
The solid line corresponds to the self-consistent RPA approximation.
The dashed line is the MF or HF result;[9]
(b) $T_c$ vs coupling constant $V$, with $t_{11},t_{20}$ fixed as in (a),
at the density where $T_c$ is maximum in (a). $V$ is in units of
$V_{AFVH} = -0.075 eV$, the coupling used in Ref.[9].
}
\end{figure}

The results for $T_c$ are
shown in Figs.3a,b compared to the MF approximation,\cite{afvh}
which in this model is
equivalent to the Hartree-Fock (HF)
approximation. The qualitative agreement is
good, and quantitatively the self-consistent approach reduces $T_c$ at
optimal doping by a factor $\sim 1.5$ still maintaining $T_c$ at a high value.
No drastic further reductions of $T_c$ are expected by adding
diagrams beyond RPA to the calculation. The reason is that in this model
the vertex correction identically $vanishes$
due to intrinsic features of the Hamiltonian Eq.(1), namely that
particles move within the same sublattice, while
the interaction is intersublattice. To understand this effect,
consider the real space representation of the
vertex correction contribution to the self-energy which is given by
$\Sigma^{vertex}({\bf r}, \tau)\propto \sum_{{\bf e_1},{\bf e_2}={\bf
\hat{x}},{\bf \hat{y}}}
G({\bf r}+{\bf e_1}, \tau)
G({\bf r}+{\bf e_1}+{\bf e_2}, -\tau)G({\bf r}+{\bf e_2}, \tau)$.
It is clear that there is always a Green's function that vanishes,
irrespective of whether ${\bf r}$ connects the same or different
sublattices. Finally, note that the values of $T_c$ shown in Fig.3a are
realistic, and the presence of an ``optimal'' density
is a consequence of a peak in the DOS of $\epsilon_{AF}({\bf k})$.\cite{afvh}

Summarizing, in this paper we studied a model for fermions
moving on a 2D square lattice with intrasublattice hopping and
attractive n.n. density-density interactions.
Using numerical and analytical techniques
we conclude that in this model (i) the two-body problem leads
to a $d_{x^2 - y^2}$-wave bound state in a natural way,
and (ii) in the dilute limit the ground state has strong $d_{x^2 - y^2}$
pairing correlations.
We have also provided evidence that the t-U-V
model actually does not show a clear signal of
d-wave SC in computational studies,
and the competition with PS prevents the analysis of its intermediate
coupling regime. Thus, we conclude that the new model discussed
here is the natural generalization to $d_{x^2 - y^2}$ superconductivity
of the attractive Hubbard model.
The new model is based on the phenomenology of the high-Tc cuprates which
near half-filling is dominated by antiferromagnetic fluctuations.
Phenomenological studies of the influence of impurities,
external fields, and other probes on
$d_{x^2 - y^2}$ superconductivity would become more accurate if
model Eq.(1) replaces  the t-U-V model.

We are grateful to P. Monthoux and F. Ortolani for many useful
discussions. We specially thank ONR for its support under
grant N00014-93-0495. We also thank the NHMFL and
MARTECH for additional support.
\medskip



\begin{references}
\bibitem{comm3}
In the 32 site cluster, level crossings at ${ |V|/t_{11} \sim 3-4}$ reduce
the signal for SC at very large $|V|$. This may be caused by
finite size effects or, more likely, by the competition with other
states like CDW or PS. From the present study
it is not clear whether the BC region of
d-wave pairs is ever stable.
\end{references}
\end{document}